\documentstyle[sprocl,twoside]{article}

\def\PRL{\em Phys. Rev. Lett.}

\newcommand{\bi}{\bibitem}
\def\be{\begin{eqnarray}}
\def\ee{\end{eqnarray}}
\newcommand{\ds}{\partial \!  \! \! /}

\begin{document}
\sloppy

\title{PARTICLE PRODUCTION IN COSMOLOGY AND 
IMAGINARY TIME METHOD
}

\author{A. D. DOLGOV}

\address{
INFN, sezione di Ferrara, Via Paradiso, 12 - 44100 Ferrara,
Italy \\
and\\
ITEP, B. Cheremushkinskaya 25, Moscow, 117259, Russia
} 

\maketitle\abstracts{ 
After brief personal recollections of the author's long-time friendship
with Misha Marinov the problem of particle production by classical time-
varying scalar field is discussed. In the quasiclassical limit the 
calculations are done by imaginary time method developed, in particular, 
in Marinov's works. The method permits to obtain simple analytical expressions
which well agree with the later found numerical solutions. The results are 
compared with perturbative calculations and it is argued that perturbation 
theory gives an upper limit for the rate of production. 
}

\vspace{0.4cm}

\tableofcontents

\newpage
\section{Personal Recollections. \label{s:psrn}}

It is difficult to write about a close friend who died so unexpectedly and
so early. Time runs fast and it seems that it was just yesterday that we 
saw each other and talked a lot on every possible subject. I really miss these 
discussions now. Last time that I saw Misha Marinov was spring 1999. 
I was visiting Weizmann Institute of Science in frameworks of Landau-Weizmann 
Program and used this opportunity to come to Haifa where Misha lived and 
worked in Technion. It was a nice sunny and fresh morning when we arrived 
with my wife Inna to Haifa railway station, where Misha met us and drove
along a beautiful road up the Carmel Mountain to his home where Lilia, his wife,
waited us with a delicious lunch. Misha was in high spirits, the four of us 
being old friends, were very glad to see each other, but slightly complained 
about, as he said, small pain in his spine.
None of us knew at that moment that it was a first sign of fast and terrible
disease.

Our friendship with Misha began, I think, in 1965 when we both were graduate 
students. We happened to be in the same plane on the way to Yerevan to
the First International Nor-Ambert School on Particle Physics. Together
with another young physicist from ITEP, Misha Terentev, we shared a small 
hut and enjoyed first in our lives international conference and charming
and hospitable town of Yerevan. 

Misha was two years older than me but I had a feeling that his knowledge
of physics, especially of mathematical physics, was at a professor level
and benefited a lot from our communications. Later we both worked in ITEP
theory group and it was always instructive and interesting to talk with him
not only about physics but practically any subject, especially
history where Misha had unusually deep and wide knowledge would it be 
ancient or modern. 

Our friendship turned into friendship between families when in 1970 we 
started to live in the same apartment building near ITEP and the distance
between our apartments was only 1-2 minute walk up or down the stairs. 
In 1979 Misha quit his position in ITEP and applied for permission to 
emigrate with the aim to live in Israel. 
Immediately life became much harder for him. It was 
difficult to find a job that could give enough money to support his family 
of four. Special rules existed at that time in the Soviet Union to 
prevent people from working without strict state control. In summer 
seasons (plus a part of spring and autumn) Misha worked as a construction 
worker, building small private houses (dachas) in the country. In winter he 
did some work for the official ``Center of Translations'' translating 
scientific papers or books from English into Russian or vice versa. However 
this kind of job was allowed only if one has another permanent place of 
work at one or other state enterprise, which Misha had not. At some stage 
they requested from him a certificate that he had such a job and since 
nothing can be presented, he was fired. So my wife, Inna, formally took 
this job and fetch for him papers to translate from the Center. Misha  
translated them, Inna presented the translated papers to the Center, 
received the money and brought it to Misha or his wife, Lilia. That's how 
it worked. 

I have to confess that I also participated in a similar deception activity, 
a few papers and books 
translated into Russian under my name were in fact translated by Misha. 
In particular, the review paper by S. Coleman 
``The magnetic monopole fifty years later''\,\cite{coleman82}
was translated for Uspekhi Fiz. Nauk, vol. 144, by Marinov but under my name.
Moreover, the editors wanted to have a short review on the activity related
to magnetic monopoles to the moment when the paper was translated, i.e.
two years after the original one had been written.
Again Misha wrote the paper and
I only signed it (honestly I also read it and liked it very much). 
So he got the money and I got the fame. Now I have to restore justice
and to change the reference\,\cite{dolgov84} into\,\cite{marinov84}.

Only in 1987 Marinovs received permission to emigrate and left for Israel. 
As we
all thought that time, emigration meant leaving for good with practically 
zero chances to see or contact each other again. However things were changing 
fast and freedom to travel abroad, unbelievable at Soviet times, came to our 
country. In 1990 Inna was able to go to Israel and for a whole month enjoyed
friendly atmosphere of Marinov's home. She and Lilia even now recall with 
mutual pleasure how nice was that time. After a couple of years Misha's life 
in Israel was successfully arranged. He got a professor position in Technion 
and was happy to be there. I remember how proudly he showed me the
campus, labs, students during my first visit to Haifa.
He enthusiastically returned to research that was interrupted for 6-7 years. 
As I can judge by what and how I learned from him, he was a very good teacher 
and did teaching with vigor and love. On the other hand, he
kept warm feelings toward ITEP, and often going in the morning
to his office in Technion he used to say, addressing Inna and Lilia, 
``Buy girls, I am going to ITEP''.

There are several fields where Misha made very important contributions, 
despite a long break in his scientific activity. But I am not going to 
describe them all, since, I think, this will be described in the Introduction 
to this volume. I will mention only two which have some relation to me.
Misha's results on application of path integral methods to complicated
quantum systems are internationally renowned and I am proud that I recommended
Maurice Jacob to publish Marinov's review on the subject in Physics 
Reports\,\cite{marinov79}. This was the last paper written by Misha, while he
was still in ITEP. Another subject where M. Marinov made a very essential
contribution together with V. Popov was electron-positron production
by an external electromagnetic field\,\cite{marinov72}. The method developed
in these works was applied to the non-perturbative 
calculations of cosmological particle production by scalar (inflaton) 
field in our paper with D. Kirilova\,\cite{dolgov90}, which is 
discussed below. 

\section{Particle Production in Cosmology; Brief Historical Review 
\label{s:hstr}}

There are two different cases of quantum particle production by 
external classical fields that are cosmologically interesting. The first
is the production by time-dependent background metric or, in other words,
by gravitational field and the second is the transformation of classical
(oscillating) inflaton field into elementary particles and the
corresponding universe (re)heating. Particle production by gravity
might be essential in the very early universe
near cosmological singularity when the strength of gravitational
field was close to the Planck value. Creation of particles by 
isotropic Friedman-Robertson-Walker (FRW) metric was pioneered by 
Parker\,\cite{parker68} and further developed
in a series of papers\,\cite{parker69,grib69,chernikov72}. Particle 
production by gravity in anisotropic cosmologies was considered in
refs.\,\cite{zeldovich70}. As argued in these papers, particle production
in anisotropic case creates anisotropic distribution of matter and back
reaction of the created matter on the metric could lead to isotropization
of the latter. Thus, in principle, the observed FRW cosmology might 
originate from a rather general initial state. More references to the 
subsequent works and detailed discussion can be found in the 
books\,\cite{books}. 

There is an important difference between particle production in 
isotropic and anisotropic cosmologies. Isotropic 
FRW metric is known to be conformally
flat, i.e. after a suitable coordinate transformation it can be reduced to 
the form:
\be
ds^2 = g_{\mu\nu} dx^\mu dx^\nu =
a^2 (r,\tau) \left( d\tau^2 - d\vec r\,^2 \right) 
\label{ds2}
\ee
From this expression follows, in particular, that FRW metric
cannot create massless particles if the latter are described by 
conformally invariant theory\,\cite{parker68}. If the particle mass, $m$, 
is non-vanishing but the interactions are conformally invariant, their 
production rate is suppressed as a power of the ratio $(m/m_{Pl})$. (Of
course, non-vanishing masses break conformal invariance.)
These statements can be easily checked in perturbation theory. The
coupling of gravity to matter fields is given by 
$\left(g^{\mu\nu}-\eta_{\mu\nu}\right) T^{\mu\nu}$,
where $\eta^{\mu\nu}$ is the Minkowsky metric tensor and
$ T^{\mu\nu}$ is the energy-momentum tensor of matter. If the
metric tensor is given by expression (\ref{ds2}), the coupling to matter
is proportional to the trace of the energy-momentum tensor that vanishes
in conformally invariant theory. 

A well known example of the theory which is conformally invariant at 
classical level (i.e. without quantum corrections)
is electrodynamics with massless charged fermions, or any other 
(possibly non-abelian) gauge 
theory describing interacting massless gauge bosons and fermions. However
quantum trace anomaly\,\cite{chanowitz73} breaks conformal invariance and
gives rise to a non-zero trace of $T_{\mu\nu}$.
In $SU(N)$ gauge theory with $N_f$
number of fermions the trace of the energy-momentum tensor of matter is
equal to:
\be
T^\mu_\mu = {\alpha \over \pi} \left( {11 N \over 3} - {2N_f \over 3}
\right) G_{\mu\nu} G^{\mu\nu}
\label{tmumu}
\ee
where $G_{\mu\nu}$ is the gauge field strength tensor. This anomaly could
strongly enhance generation of electromagnetic field (or any other gauge 
fields) in the early universe\,\cite{dolgov80}.

Another simple and important theory of a free massless scalar field 
$\phi$ is not conformally invariant even at the classical level 
if $\phi$ is minimally coupled to 
gravity (that is through covariant derivatives only). The energy-momentum
tensor of such field is given by the expression:
\be
T_{\mu\nu} (\phi) = (1/2)\partial_\mu \phi \, \partial_\nu \phi - 
(1/4) g_{\mu\nu}  \partial_\alpha \phi\, \partial^\alpha \phi
\label{tmunuphi}
\ee
and its trace $T^\mu_\mu = -(1/2)\partial_\alpha \phi\,\partial^\alpha \phi$
is generally  non-vanishing. Conformal invariance can be restored if one
adds to the free Lagrangian the nonminimal coupling to gravity in the
form $R \phi^2 /12$ (see e.g. refs.~\cite{books}). However it would be
better not to restore it because generation of primordial density
perturbations at inflation\,\cite{guth82}, which serve as seeds for large 
scale structure formation, is possible only for non-conformal fields.

Another realistic example of conformally non-invariant theory with massless
fields is gravity itself. It was shown that gravitational waves are not
conformally invariant in the standard General Relativity\,\cite{grishchuk75}.
This explains efficient production of gravitational waves during inflationary
stage\,\cite{starobinsky79}. 

A renewed interest to gravitational particle production arose
in connection with a possible explanation of the observed ultra-high 
energy cosmic rays by heavy particle decays\,\cite{berezinsky97}. 
There are two competing mechanisms of creation of such particles in
cosmology: by background metric
and by inflaton field. The former was considered in refs.~\cite{kuzmin98}
(for a review see\,\cite{kuzmin99}), while particle production by inflaton
will be discussed below.

In the earlier papers\,\cite{dolgov82} the
universe (re)heating at the final stage of inflation through particle
production by the oscillating inflaton field was treated in a simplified 
perturbation theory approximation. First non-perturbative treatment
was performed in two papers\,\cite{dolgov90,traschen90}.
In what follows we concentrate on the approach of ref.\cite{dolgov90} 
where the imaginary time method was used.
In both papers\,\cite{dolgov90,traschen90} a possibility of 
parametric resonance enhancement of particle production rate, noticed long 
ago\,\cite{narozhnyi73}, was mentioned. However, it was argued in the first 
of them that the resonance was not effective because the produced particles 
were quickly removed from the resonance band by the cosmological expansion 
and elastic scattering on the background. A more careful analysis of the
subsequent paper\,\cite{traschen90} showed that under certain condition 
expansion might be irrelevant and did not destroy the resonance. In this 
case a strong amplification of the production probability and much faster
process of post-inflationary (re)heating could be expected. The issue of
the parametric resonance (re)heating attracted great attention after the 
paper\,\cite{kofman94}, and now the number of published papers on the subject
is measured by a few hundreds. However, a review of this activity is 
outside the scope of the present paper and below we will confine ourselves
to the problem of fermion production by a time dependent scalar field
where parametric resonance is not effective.

Concerning production of fermions, there is a contradiction in 
the literature between the paper\,\cite{dolgov90}, where non-perturbative 
production of fermions
was pioneered, and the subsequent ones. While in the paper\,\cite{dolgov90} 
was stated that fermion production is always the strongest in perturbation 
theory regime, and in the opposite, quasiclassical
limit the production is noticeably weaker, in subsequent works was
argued that in non-perturbative regime fermion production was strongly
enhanced so that it could even compete with resonant boson production.
Calculations in ref.~\cite{dolgov90} have been performed by imaginary time
method, while other works either used numerical calculations or some
approximate analytical estimates. I will argue in what follows that
there is practically no difference between the results of all calculations,
earlier and later ones, but the difference is in the interpretation of the 
results and fermion production by the inflaton is always weak, weaker than
that found in perturbation theory.

\section{Particle Production in Perturbation Theory \label{s:pert}}

Let us start from consideration of production in the case when perturbation
theory is applicable and calculations are straightforward and
simple. In this section
we will neglect the universe expansion and assume that the external scalar
field periodically changes with time according to:
\be
\phi (t) = \phi_0 \cos \omega t
\label{phioft}
\ee
Here $\phi_0$ is the amplitude of the field, it can be slowly varying function
of time, and the frequency of oscillations $\omega$ coincides with the
mass of $\phi$ if the latter lives in the harmonic potential
$U(\phi) = m^2_\phi \phi^2/2$. 

We assume that $\phi$ is coupled to fermions through the Yukawa interaction:
\be
{\cal L}_{\psi} = \bar \psi \left( i\ds +m_0 \right) \psi \,+\,
g \phi \bar \psi \psi
\label{lpsi}
\ee
Perturbation theory would be valid if the coupling constant is small,
$g\ll 1$, which is well fulfilled for the inflaton field, and if the
fermion mass is smaller then the mass of the inflaton,
$m_\phi = \omega$. The last condition may not be true even if $m_0 < m_\phi$
because the interaction with $\phi$ introduces effective time-dependent mass
\be
m_1 (t) = g \phi_0 \cos \omega t
\label{m1oft}
\ee
and for a large amplitude $\phi_0$
the latter may be large in comparison with $\omega$ for most of the 
oscillation period, except for a small part, when $\cos \omega t$ is close 
to zero. In this case perturbation theory is invalid. 

It is practically evident, even without calculations, that in
perturbative case the rate of particle production is equal to 
the width of the decay of the scalar boson $\phi$ into a pair of fermions:
\be
\dot n_\psi /n_\phi = \Gamma_\phi = g^2 \omega /8\pi
\label{dotn/n}
\ee
where $n_{\psi ,\phi}$ are the number densities of  $\psi$ and $\phi$
particles per unit volume respectively
and we assumed for simplicity that the fermion mass $m_0 =0$ (it is 
straightforward to lift this restriction).

Still to make comparison with subsequent non-perturbative calculations we
will sketch below the derivation of this result. According to general rules of 
quantum field theory the amplitude of production of a pair of particles with 
momenta $\vec p_1$ and $\vec p_2$ by an
external time-dependent field $\phi(t)$ in first order in perturbation
theory is given by 
\be
A(\vec p_1,\vec p_2) = g\int d^4 x \phi(t) 
\langle \vec p_1,\vec p_2 | \bar \psi (x) \psi (x) | vac \rangle 
\label{apt}
\ee
where the state $\langle \vec p_1,\vec p_2 |$ is produced by action on 
vacuum of the creation
operators in the standard second-quantized decomposition of Dirac operators
$\psi $ and $\bar \psi$:
\be
\psi (x) = \sum_s \int{d^3k \over (2 \pi)^{3}} 
[u_k^s b_k^s e^{-ik \cdot x} + v_k^s d_k^{s{\dagger} }
e^{ik \cdot x}] 
\label{psiofx}
\ee
where $b_k^s$ and $d_k^{s{\dagger}}$ are respectively annihilation and 
creation operators for particles and
antiparticles with momentum $k$ and spin $s$.

After the usual anti-commutation algebra we will arrive to the integral
\be
\int d^3 k d^3 k' \delta (\vec k - \vec p_1)  \delta (\vec k'- \vec p_2)
e^{i(E+E')t - i(\vec k+ \vec k') \vec x } 
\label{intdp}
\ee
The integral can be trivially taken and substituted into the integral
over $d^3x dt$ (\ref{apt}). Integration over $d^3 x$ gives
$\delta (\vec p_1 +\vec p_2)$ and we are left with the Fourier
transform:
\be
A(\vec p_1,\vec p_2) \sim g^2 \delta^{(3)} (\vec p_1 +\vec p_2)
\int dt \phi (t) e^{i(E_1+E_2)t}
\label{adelta}
\ee
(for details and more rigorous consideration in terms of Bogolyubov
coefficients see e.g. appendix A in ref.~\cite{dolgov96}).

The probability of particle production is proportional to 
$ |A(\vec p_1,\vec p_2) |^2$ and contains the square of momentum
delta-function. The latter is treated in the standard way,
\be
[\delta (\vec p_1 +\vec p_2)]^2 = 2\pi V \delta (\vec p_1 +\vec p_2)
\label{delta2}
\ee
where $V$ is the total space volume. The origin of the volume factor is 
evident: since the external field is space-point independent, so is the 
probability of production per unit volume and the total probability 
is proportional to the total volume. 

Similar situation is realized for the time dependence in the
case of periodic external
fields, if one neglects back reaction of the produced particles on the
field evolution and on the probability of production. The former can be taken
into account by a (slow) decrease of the field amplitude $\phi_0(t)$, while 
the latter is determined by the statistics of the produced particles: the
probability of boson production is proportional to the phase space density 
of already produced bosons, $(1+f_k)$, while the probability of fermion 
production is inhibited by the factor $(1-f_k)$. This back reaction effect 
is absent for Boltzmann statistics, which we will mostly assume in what
follows.
Thus, for a periodic external field one would expect that the probability
of production is proportional to the total time interval, during which the
external field was operating. In the idealistic case of 
$\phi \sim \exp (i\omega t)$, its Fourier transform gives 
$\delta (2E -\omega)$ and the square of the latter is, as above,
$t_{tot} \delta (2E -\omega)$. The second factor ensures energy conservation
and is infinitely large for $E=\omega/2$. It means that the phase space
density of the produced particles becomes very large after period of time
when the energy conservation is approximately established. One can check
that this time is much shorter than $1/\Gamma$ (where $\Gamma$ is the 
perturbative decay rate) but still the time of transition of energy from the
inflaton field to the produced fermions is given by $1/\Gamma$. This fact
is commonly agreed upon in the case of perturbative production. The 
statements in the literature that in non-perturbative regime fermion
production could be very strong is possibly related to this trivial rise
of the occupation numbers and does not mean that fermion production
can compete with production of bosons (see below).

In the case when external field operates during a finite period of time,
starting e.g. from $t=0$,
or if one is interested in the number of produced particles at the running
moment $t$, the integral in expression (\ref{adelta}) should be taken in
the limits $(0,t)$ and for the particular case of 
$\phi = \phi_0 \cos \omega t$ one obtains:
\be
I(t; E, \omega) &\equiv& \int_0^t dt\, e^{2iEt}  \cos \omega t \nonumber \\ 
&=&e^{i(E-\omega/2) t}\left[ {\sin (E - \omega/2) t \over 2E -\omega }
+ e^{i\omega t}\,{\sin (E + \omega/2) t \over 2E +\omega}\right]
\label{int0t}
\ee

For $E$ close to $\omega/2$ the first term dominates and 
the number of produced fermions rises as $t^2$ till $t \sim 1/|2E-\omega|$.
At larger times it oscillates. The same phenomena was found in 
non-perturbative calculations. Indeed, the phase space number density 
of the produced particles (we use this term interchangeably with the 
``occupation number'') is given by
\be
f_p = g^2 \phi_0^2 \mid I (t; E,\omega) \mid^2
\label{fpptbl}
\ee
As we have argued above, usually one has 
$ \mid I (t; E,\omega) \mid^2 = 2\pi t\,\delta (2E-\omega)$. In this case the
number density of the produced particles as a function of time is given by:
\be
n(t) =\int {d^3 p \over (2\pi)^3} f_p = 
{g^2 \omega \over 8\pi}\,\,\phi_0^2\omega t = \Gamma n_\phi t
\label{noftbl}
\ee
where $n_\phi = \phi_0^2 \omega$ is the number density of $\phi$-bosons
and $\Gamma$, given by eq.~(\ref{dotn/n}), is their decay width. 

A detailed explanation of the discussed phenomena can be found in 
textbooks on quantum mechanics in the section where perturbation theory 
for time dependent potential is presented, see e.g.\cite{landau}.

Returning to the occupation number (\ref{fpptbl}) we see that 
for $(\omega- 2E)t <1$ it evolves as $f_p \approx g^2\phi_0^2 t^2$ and
reaches unity for $t=t_1 = 1/g\phi_0$. This is much earlier than 
$t_d = 1/\Gamma$ which is the characteristic decay time of $\phi (t)$:
\be
t_d /t_1 = (8\pi /g)\,(\phi_0 /\omega) 
\label{tdt1}
\ee
Formally taken this ratio may reach the value $10^8-10^9$.
This is an explanation of statement that fermions could be very quickly 
produced by inflaton. On the other hand, though some fermionic bands
(approximately satisfying energy conservation)  
might be quickly populated, the total transfer of energy from the
inflaton to the produced particles is determined by the total decay rate
and is much slower. Roughly speaking $f_p = 1$ corresponds to production
of only one pair of fermions and, of course, the energy of this pair
is negligible in comparison with the total energy accumulated in the
classical field $\phi (t)$.

Perturbation theory is not applicable if the effective mass of 
fermions $m_{eff}= (m_0 + g\phi_0)$ is larger than the frequency of the 
oscillations of the scalar field. For example, the probability of pair
production by two-quanta process, when the energy of each produced
fermion would be equal to $\omega$, is related to one-quantum process,
when $E=\omega/2$, as $W_2/W_1 \sim (g\phi_0 /\omega)^2$. It is still
possible that $\phi_0/g\omega \gg 1$, while $g\phi_0/\omega <1$, so that
perturbation theory is reliable and the relation $t_d/t_1 \gg 1$ still 
holds. However in many practically interesting cases 
$g\phi_0/\omega >1$ and in this range of parameters the
result obtained above can serve only for the purpose of
illustration and for more precise statements 
we have to go beyond perturbation theory. This will be
done in the following section by the imaginary time 
method\,\cite{nikishov69,popov71,marinov72}. (For recent applications of 
this method and a more complete list of references see\,\cite{ringwald01}.)
Qualitatively clear that 
non-perturbative effects could only diminish the rate of particle 
production because the non-perturbative calculations take into 
account non-vanishing and large value of the effective mass of the produced 
particles and this leads to a smaller rate of the production in comparison
with the case when the interaction is taken in the form $g\phi\bar\psi\psi$
but its contribution into fermion effective mass is neglected. As we see
below, the suppression of the production rate in nonperturbative
regime\,\cite{dolgov90} in comparison with perturbation theory is given 
by the factor $(\omega/g\phi)^{1/2}$ in qualitative agreement with these
simple arguments.

Effects of quantum statistics were neglected above, and thus the results
obtained are valid only if $f_p <1$. The corresponding corrections can be
approximately introduced by multiplication of the r.h.s. of 
eq.~(\ref{fpptbl}) by the factor $(1\pm f_p)$ and correspondingly
$f_p^{(f,b)} = g^2 \phi_0^2|I|^2 / (1 \pm  g^2 \phi_0^2 |I|^2)$, 
where the signs
$''\pm''$  refer for fermions and bosons respectively. One sees that 
the production of fermions effectively stops (as one should expect) when
$f_p^{(f)} \sim 1$, while production of bosons tends to infinity. Presumably 
a more accurate treatment would not allow bosons to reach infinitely large
density in a finite time but the message is clear, the production of bosons
becomes explosive in perturbation theory with characteristic time of the
order of $t_1 = 1/(g\phi_0)$ and all the energy of the inflaton would go 
into that of the produced bosons during approximately this time. There are
several effects that can weaken this conclusion. One is a possible 
inapplicability of perturbation theory for a large $g\phi_0/\omega$. This 
effect qualitatively acts in the same way as in 
fermionic case discussed above. 
Still, even if the  $g\phi_0/\omega>>1$ the effect of explosive production
of bosons would survive due to parametric resonance in equation of motions
for the produced modes\,\cite{dolgov90,traschen90,kofman94}. Another two 
effects that could diminish the production are the cosmological red-shift 
of momenta of the produced particles and their scattering on other 
particles in the background. Both would push the produced particles away 
from the resonance band and
could significantly slow down the production in the case of narrow 
resonance\,\cite{dolgov90}, while in the case of wide resonance the effect 
survives\,\cite{traschen90,kofman94}. 

On the other hand, both red-shift and scattering of the produced fermions
back react on their production in exactly opposite (to bosons) way. 
These phenomena ``cleans'' the occupied zone and allows for 
production of more fermions.

\section{Quasiclassical Limit; Imaginary Time Method. \label{s:imt}}
\subsection{Small Mass Case. \label{ss:small}}

Usually non-perturbative calculations are not simple but in the case
that we are considering there is a fortunate circumstance that in the
anti-perturbative limit quasiclassical approximation works pretty well.
The latter can be efficiently treated by the imaginary time 
method\,\cite{nikishov69,popov71,marinov72}. Below we will essentially 
repeat the paper\,\cite{dolgov90} correcting some
typos and algebraic errors, though the basic results of the paper remain
intact.

The coupling of $\phi(t)$ to the produced particles is equivalent to 
prescription of the time dependent mass to the latter,
$m(t) = m_0 + g\phi(t)$. The classical Lagrange function for 
a relativistic particle with such a mass has the form
\be
L_{cl} = - m(t) \left( 1 - \vec V^2\right)^{1/2}
\label{lcl}
\ee
where $\vec V$ is the particle velocity.
The corresponding Hamiltonian is 
\be
{\cal H} = \left[ p^2 + m^2 (t)\right]^{1/2} \equiv \Omega (t)
\label{ham}
\ee
The quantization of this system can be achieved by the path integral
method. The Green's function of the quantum particle has the form
(see e.g.~\cite{marinov79}):
\be
G(\vec x_f,t_f; \vec x_i, t_i) = \int D\vec p \, D\vec x 
\exp\left[ i\int_{t_i}^{t_f} dt \left(\vec p\,\dot {\vec x} - {\cal H}
\right)\right].
\label{gxtint}
\ee
The functional integral in this case can be easily taken, giving:
\be
G(\vec x_f,t_f; \vec x_i, t_i) = \int {d^3 p \over (2\pi)^3}
\exp \left[ i\vec p\,\left(\vec x_f-\vec x_i\right) 
-i\int_{t_i}^{t_f} dt \Omega (t) \right]
\label{gxt}
\ee

According to the general rules of quantum mechanics the amplitude of the
transition from the state given by the initial wave function $\Psi_i$ into
that given by $\Psi_f$ is equal to 
\be
A(\vec p_1,\vec p_2) = \int d^3 x_i d^3 x_f \Psi^*_f (x_f) 
G(\vec x_f,t_f; \vec x_i, t_i) \Psi_i (x_i).
\label{ap1p2}
\ee
where for $\Psi_{i,f}$ plane waves are usually substituted.

If we want to obtain the amplitude of creation of a pair of particles
the contour of integration over time should be shifted into complex 
$t$-plane in such a way that it goes around the branching point of the
energy $\Omega$ in the direction of changing the sign of energy from 
negative to positive one. This corresponds to transition from the lower
continuum of the Dirac sea to the upper one, i.e. to pair creation.
Thus we find:
\be
A(\vec p_1,\vec p_2) = \left( 2\pi \right)^3 \delta \left( \vec p_1  +
\vec p_2 \right) \exp \left[ -i \int_{C(t_i,t_f)} dt\,\, \Omega (t)
\right],
\label{apair}
\ee
where the  contour $C(t_i,t_f)$ starts at $t=t_i$ and ends at $t=t_f$
and turns around the branching point of $\Omega$ in the way specified 
above. 

The position of the branching points $t_b = t' +it'' $ can be found 
from the equation:
\be
p^2 + \left( m_0 + g \phi_0 \cos \omega t \right)^2 =0.
\label{brpoint}
\ee 
Correspondingly
\be
m_0 + g\phi_0 \left( \cos \tau'\cosh \tau'' -i \sin \tau' \sinh \tau'' 
\right) = \pm i p,
\label{tt}
\ee
where $\tau = \omega t$.
In what follows we assume that $m_0 = 0$ and it will grossly simplify
technical details. In this limit $\tau' = \pi/2 + n\pi$ and 
$\sinh \tau'' = \pm (p/g\phi_0)$.

The integral along the cut $\tau = \tau' +i\eta$ is real and, according
to our prescription, negative. It gives exponential suppression factor
for the production probability,
$W \sim \exp (-2Q)$, with
\be
Q =(2/\omega) \int_0^{\tau''} d\eta \left( p^2 - g^2\phi_0^2\,\sinh^2 \eta 
\right)^{1/2}.
\label{Q}
\ee
This integral can be expressed through complete elliptic 
functions\,\cite{grad}:
\be
Q = {2\sqrt{p^2 + m_1^2} \over \omega}\, 
\left[ K\left( \beta\right) - E(\beta) \right] 
\label{qell}
\ee
where 
\be
m_1 = g\phi_0,\,\,\,{\rm and}\,\,\,
\beta = p/\sqrt{p^2 + m_1^2}.
\label{beta}
\ee
For small $\beta$ these
functions can be expanded as $K(\beta) \approx (\pi/2) (1+\beta^2 /4)$
and $E(\beta) \approx (\pi/2) (1-\beta^2 /4)$, so that
$Q\approx (\pi/2) (p^2/ \omega \, m_1 ) $.

The total production amplitude is equal to the sum of 
expressions~(\ref{apair}) with all the contours encircling the proper 
branch points between $t_i$ and $t_f$. Since the integrals along imaginary 
direction $id\eta$ are all real and have the same value for all branch 
points, their
contribution to the amplitude gives the common factor $\exp (-Q)$.
The integrals over real time axis corresponding to different contours
$C$ around neighboring branch points differ by the phase factor
$A_{n+2}/A_n =\exp (2i\alpha)$, because the energy changes sign after
the integration contour turns around branch points. The absence of the
contribution from the nearest cut is related to the particle statistics and
is discussed e.g. in ref.\cite{popov71,marinov72}. 
The phase $\alpha$ is given by:
\be
\alpha = \int_0^{2\pi} dt \,\sqrt{p^2 +m_1^2 \cos^2 \omega t}  
= {4\,\sqrt{p^2 + m_1^2} \over \omega}
\, E\left(\sqrt{1-\beta^2}\right)
\label{alpha}
\ee
All this is true if the free fermion mass is vanishing, $m_0 =0$,
otherwise equations become significantly more complicated.
In the limit of small $\beta$ we find\,\cite{grad}:
\be
E(\sqrt{1-\beta^2})\approx [1 + (\beta^2/2) (\ln (4/\beta)-1/2)]
\label{E}
\ee
while for $\beta$ close to 1 the necessary expressions are presented after
eq.~(\ref{beta}) with the interchange $\beta^2 \leftrightarrow (1-\beta^2)$. 

Summing over all branch points we obtain:
\be
A(\vec p_1,\vec p_2) = \left( 2\pi \right)^3 \delta \left( \vec p_1  +
\vec p_2 \right)\,\exp {\left(- Q + i\alpha\right)}\,
{\sin (N  \alpha ) -1 \over \sin ( \alpha ) -1}
\label{atot}
\ee
where $N$ is the total number of branch points included in the amplitude;
it is approximately equal to the total time in units $1/\omega$ during 
which the particles are produced, $N= {\rm Integer}[ (t_f-t_i)/\omega]$.
The last factor reminds that coming from the integration over time in
perturbation theory discussed in sec.~\ref{s:pert} and in fact its physical
nature is the same. For very large $N$, formally for $N\rightarrow\infty$
it tends to
\be
{\sin (N\alpha ) \over \sin \alpha }\rightarrow 
\pi \sum_j \delta \left( \alpha -\pi j \right)
\label{delofal}
\ee 
These delta-functions impose energy conservation for the production of
pair of particles by $j$ quanta of the field $\phi$. Note that in contrast
to the lowest order perturbation theory, when only a single quanta production
is taken into account, the expression (\ref{atot}) includes production of a
pair by many quanta of the field $\phi$. For example, in the limit of high
momenta of the produced particles these delta-functions are reduced to
$\delta (2p -j\omega)$, the same as in perturbation theory for j-quanta
production.

Treating again, as in sec~\ref{s:pert}, the square of delta-function as
a product of the single delta-function and $\delta (0) = \pi N$ with
$N$ expressed through the total time $t$, during which the particles have 
been produced, as the integer part of $t\omega $, 
we find the following expression for the rate of production per unit time 
and unit volume\,\cite{dolgov90}:
\be
\dot n =\pi\,\omega \sum_j \int {d^3 p \over (2\pi)^3 } \, \exp (-2Q)\, 
\delta \left( \alpha - \pi j\right)
\label{dotn}
\ee
In the limit of $m_1 \gg \omega$ one obtains:
\be
Q &\approx& {\pi\over 2}\,{p^2 \over \omega\, m_1}\\
\alpha &\approx& {4m_1 \over \omega}\left[ 1 + {p^2 \over 2m^2_1}
\left( \ln {4m_1 \over p} +1\right)\right],
\label{Qalpha}
\ee
and hence
\be
\dot n = {1\over 2\pi}\,\sum_{j_m} \exp\left[-{\pi^2\left(j - 
(4m_1/\pi\omega) \right) 
\over \ln (4m_1/p_j) +1} \right] {\omega^2 m_1 p_j \over
\ln (4m_1 /p_j) +1/2}
\label{dotnpj}
\ee
Here summation starts from the minimum integer value
$j_m \geq (4m_1/\pi\omega)$ and $p_j$ is determined from the equation
$\alpha = j\pi$, i.e.
\be 
p_j^2 \approx (\pi m_1 \omega/2)(j-4m_1/\pi\omega) /[\ln(4m_1/p_j) +1]
\label{pj2}
\ee
A rough estimate gives $\dot n \sim \omega^{5/2} m_1^{3/2}$. Correspondingly
the characteristic rate of the inflaton decay in the quasiclassical 
approximation is given by 
\be
\Gamma_{q} = \dot n/n_\phi =\dot n /(\omega \phi^2_0) 
\sim \Gamma \left( \omega /m_1 \right)^{1/2}
\label{gammaq}
\ee
where $\Gamma$ is the decay rate in perturbation theory~(\ref{dotn/n}).
One sees that in the quasiclassical limit the decay rate is suppressed
in comparison with the formal result of perturbation theory 
as a square root of the ratio of the oscillation frequency to the amplitude
of the scalar field. This suppression can be understood as 
follows\,\cite{dolgov90}. Most of the time the instant value of the field
$\phi(t)$ and the effective mass of the fermions, 
$m_{eff}=g\phi_0\,\cos \omega t$ are large in comparison with 
the oscillation
frequency. As is well known (see also sec. \ref{ss:large} below) the
probability of particle production in this case is exponentially suppressed.
However, when $\cos \omega t$ is very close to zero the effective mass of the
produced particles would be smaller than $\omega$ and they are essentially
produced at this short time moments. This results in a much milder
suppression of the production, not exponential but
only as $(\omega/g\phi_0)^{1/2}$.

For the case of finite and not too big $N$ we will see that, according to
the calculations of reference\,\cite{dolgov90} presented above,
the occupation number $f_p$ would reach unity in a much shorter time 
than $1/\Gamma_q$. This result was rediscovered later in the 
papers\,\cite{baacke98,green99} by numerical calculations and 
reconfirmed by analytical 
methods in ref.~\cite{peloso00}. However, as it has been already argued, 
this does not mean that non-perturbative production of fermions is 
strong, it is always weaker than the perturbative one. 

The calculations presented above do not include the effects of quantum 
statistics, so strictly speaking, they are valid for ``boltzons''. 
Thus, they present an upper bound for the
production of fermions. In the fermionic case, the production would stop 
when the occupation number, $f_p$, approaches unity, while production of
``boltzons'' would go unabated. However, if the particles from the occupied 
Fermi band are quickly removed by scattering or red-shift (as we discussed 
above) the production of fermions would go essentially with the same rate as
production of ``boltzons''. 

For a finite number of oscillations $N$ the occupation number of the produced
particles is equal to (see eq. (\ref{atot})):
\be
f_p (N) = \exp (-2Q)\,\left( {\sin (N  \alpha ) -1 
\over \sin ( \alpha ) -1} \right)^2
\label{fpnonpt}
\ee
The last factor is rather similar to that in eq.~(\ref{int0t}). This is an
oscillating function of $N$. For $\alpha= \pi (1-\epsilon)$ with a small
$\epsilon$ it rises roughly as $N^2$ during $N= 1/(2\epsilon)$ oscillations.
The occupation number increases with time discontinuously as a series of 
discrete jumps as time $t/\omega$ reaches integer values. During this
stage $f_p$ may quickly rise with the speed much faster then the rate
$\Gamma_q$~(\ref{gammaq}) in complete analogy with the perturbative
case considered in sec.~\ref{s:pert}. However, as we have already stressed,
this does not mean that the production of fermions goes faster than 
in perturbation theory.
 
After this period of increase, $f_p$ starts to go down and approaches zero 
at $N_0\approx 1/\epsilon$. This oscillating behavior of the number of produced 
particles was noticed long ago in the problem of $e^+e^-$-pair creation
by periodic electric field (for the list of references see e.g. the 
book by Grib et al in ref.~\cite{books}). Thus it looks as though particles
are produced by the field and after a while they all are absorbed back. This
behavior is difficult to digest. Note that it is absent if time is very large,
tending to infinity, as is discussed above. In this case the energy 
conservation is strictly imposed by the delta-function, $\alpha = \pi n$
(where $n$ is an integer), or in other words $\epsilon =0$ and 
$N_0 \rightarrow \infty$.

Possibly this mysterious phenomenon of re-absorption of the produced particles
is related to the fact that during finite time the external field
$\phi(t)$ does not disappear and the particle vacuum is not well defined 
over this time dependent background. To resolve the ambiguity  one may calculate 
the transition of energy from the time-varying field $\phi (t)$ into other 
quantum states which are not necessarily determined in terms of particles.
Energy density, in contrast to the particle number density, can
be unambiguously defined in terms of local fields operators and does not
suffer from any ambiguity related to the non-local character of the latter.
The energy density of the quantum field $\psi$, defined as the
expectation value of the time-time component of its energy-momentum operator,
may also exhibit the oscillating behavior described above
but the correct interpretation is 
possibly not production of $\psi$-particles but some excitation ("classical"?)
of the (fermion) field $\psi$ coupled to $\phi(t)$.

\subsection{Large Mass Case. \label{ss:large}}

Let us now consider the case when the fermion mass $m_0$ is large in
comparison with the oscillation frequency $\omega$ and 
with the amplitude of
the oscillations, $m_0 \gg g\phi_0$, so that the total effective fermion 
mass, $m_{tot} = m_0 + g\phi_0 \cos \omega t$ never vanishes and
always large. The 
calculations for this case have been only done in ref.~\cite{dolgov90} and
we will reproduce them here. To be more precise we will reproduce only 
imaginary time part, while in ref.~\cite{dolgov90} the method of 
Bogolyubov coefficients was used as well.

Following this paper we will consider production of bosons. It will be
technically simpler allowing to make all calculations analytically, but
qualitatively the same results should be valid also for fermions, because
for a large $m_0$ the production is weak and the occupation numbers remain
small. We assume that the effective mass has the form
\be
m^2(t) = m_0^2 + g^2 \phi_0^2 \cos^2 \omega t
\label{m2oft}
\ee
This case is realized if the interaction of the inflaton field with the
produced particles ($\chi$-bosons) has the form $g^2|\chi^2| \phi^2$.
The probability of production can be found from the expressions of the 
previous subsection by the substitution $p^2 \rightarrow  p^2 +m_0^2$.
In particular, the exponential damping factor is given, 
instead of~(\ref{qell}), by:
\be
Q' = {2\sqrt{p^2 +m_0^2 + m_1^2} \over \omega}\, 
\left[ K\left( \beta'\right) - E(\beta') \right] 
\label{qell'}
\ee
where 
\be
(\beta')^2 \equiv 1- u^2 = 1 - {m_1^2 \over m_0^2 +m_1^2 + p^2}
\label{beta'2}
\ee
and the complete elliptic integrals in the case of small $k$ are expanded 
as\,\cite{grad}:
\be
K(\beta') &\approx& \ln {4\over u} + {u^2\over 4}\, \left(\ln  {4\over u}
-1\right)\nonumber\\
E(\beta') &\approx& 1 +  {u^2\over 2}\, \left(\ln  {4\over u} -{1\over 2}
\right)
\label{KE}
\ee
The phase difference over the period of oscillations is now given by:
\be
\alpha' &=&  {4\,\sqrt{p^2 + m_0^2+ m_1^2} \over \omega}
\, E\left(\sqrt{{m_1^2 \over m_1^2+m_0^2+p^2}}\right) \nonumber \\
&\approx&  {2\pi\,\sqrt{p^2 + m_0^2+ m_1^2} \over \omega}
\left[ 1+ {m_1^2 \over 4(m_1^2+m_0^2+p^2)}\right]
\label{alpha'}
\ee

We can repeat the same calculations as in the previous subsection to find
the occupation number and the number density of the produced particles.
The production probability is now exponentially suppressed, as
$\exp \{{-2 \sqrt{m_0^2+m_1^2}\ln [16(m_0^2+m_1^2)/m_1^2] /\omega}\}$.
For a sufficiently large ratio $m_0/\omega$ the production would be
very weak, all occupation numbers would be small in comparison with unity
and bosons and fermions would be equally poorly produced.

\section{Back Reaction and Cosmological Expansion Effects. \label{s:evol}}.

Now we briefly comment on applicability of the results discussed
above to realistic case of universe (re)heating after inflation.
We have neglected universe expansion and damping of the field $\phi$
due to energy transfer to the produced particles. The effect of expansion
can be easily taken into account in conformal coordinates where the
metric takes the form~(\ref{ds2}) with space point independent
cosmological scale factor $a(\tau)$. 
Under transformation to conformal coordinates and simultaneous redefinition
of the gravitational, scalar, and fermionic fields respectively as
$g_{\mu\nu} \rightarrow a^2 g_{\mu\nu}$, $\phi \rightarrow \phi/a$, 
and $\psi \rightarrow \psi /a^{3/2}$, the mode equation for the scalar
field takes the form:
\be
\phi_k'' +(k^2 + m^2 a^2 - a''/a ) \phi_k = 0,
\label{phi''}
\ee
where the derivatives are taken with respect to conformal time and $k$ is 
comoving momentum. The presence
of the term $a''/a$ demonstrates breaking of conformal invariance even for
massless scalar field, as has been already mentioned in sec.~\ref{s:hstr}. 
All masses enter equation of motion in the combination $ma$, so mass
terms explicitly break conformal invariance. The interactions of
the types $g\phi \bar \psi \psi$, $\lambda \phi^4$ and
$f\phi^2 \chi^* \chi$ are invariant with respect to the transformation
of the fields specified above (note that the presence of the 
$\sqrt{det[g_\mu\nu]}$ in the action integral gives the necessary 
factor $a^4$ to ensure this invariance).

The expressions for the scale factors through conformal time in three
most interesting cosmologies are the following:
\be
a&\sim& e^{Ht} = - {1/ H\tau}, \,\,\, {\rm De Sitter\,\,\,universe,\,\,\,
inflation},\nonumber\\
a&\sim& t^{1/2} \sim \tau, \,\,\, {\rm radiation\,\,\,dominance},\nonumber\\
a&\sim& t^{2/3} \sim \tau^2, \,\,\, {\rm matter\,\,\,dominance}.
\label{expregm}
\ee
In particular, in the radiation dominated universe with conformally
invariant interactions, scalar field is conformally invariant but this is
not true for other expansion regimes. Correspondingly, particles production by 
massless scalar field with the self-potential $\lambda \phi^4$ can be reduced
to the flat space case discussed in the previous section. The difference
between the potentials of $\phi$ in these two cases, $\omega^2 \phi^2$
and $\lambda \phi^4$, is not essential and the obtained above results can be
easily translated to the $\lambda \phi^4/4$ potential. Indeed, the equation
of motion of spatially homogeneous field $\phi$ in flat space-time 
(in conformal coordinates) has the
form:
\be
\phi'' +\lambda \phi^3 = 0
\label{ddotphi}
\ee
This equation is solved in terms of Jacobi elliptic functions\,\cite{grad}:
\be
\phi (\tau) &=& \phi_0 \,{\rm cn} \left( \sqrt{2\lambda}\phi_0\tau; 
\sqrt2 \right)
\nonumber \\
&=& {2\sqrt{2} \pi \over \kappa} \sum_{n=1} {\exp [-\pi(n-1/2)] \over
1 + \exp [-\pi (2n -1)]}\,
\cos \left[\left(2n-1\right){\pi \sqrt{2\lambda}\phi_0 \tau 
\over 2\kappa } \right]
\label{cnoft}
\ee
where $\kappa = \Gamma^2 (1/4) /4\sqrt{\pi}$. The expansion is well 
approximated by the first term and particle production rate can be estimated 
using results of the previous section. Significant deviations from those 
results can be expected only in the case of heavy particle production when 
higher frequency terms in expansion~(\ref{cnoft}) may compete with the
exponentially suppressed contribution coming from 
lower terms  (see eq.~(\ref{qell'})).

It should be repeated, however, that these results are true only for radiation
dominated regime of expansion. For other cosmologies the term $a''/a$ in
eq.~(\ref{phi''}) is non-vanishing and must be taken into account.

Another effect, in addition to expansion, that results in a decrease of 
the amplitude of the field $\phi(t)$, is back reaction of the particle
production. Energy that is transferred to the produced particles
is taken from the field $\phi$ so the energy density of the latter should
become smaller. For harmonic oscillations (in the case of the potential
$\omega^2 \phi^2$) only the amplitude of the field diminishes, while 
frequency remains the same. For quartic potential both the frequency and the
amplitude of oscillations go down, as one can see from eq.~(\ref{cnoft}) with
$\phi_0 (t)$.

In the case of quickly oscillating field the effect can be easily estimated 
in adiabatic approximation. One has to solve the equation for energy balance
in expanding background:
\be
\dot \rho = -3H\, \left( \rho + P \right)
\label{dotrho}
\ee
where $\rho$ and $P$ are respectively energy and pressure densities of the 
field $\phi$ and the produced particles. For the former the solution of the
standard equation of motion without interactions should be substituted with
the effect of production included in a slow decrease of the amplitude 
$\phi_0$. 

More accurate consideration demands using equation of motion modified by the
production process. Usually this is described by the introduction into
equation of motion, in addition to Hubble friction,  
the ``production friction term'':
\be
\ddot \phi + 3H \dot \phi + U'(\phi) = -\Gamma \dot \phi.
\label{dotphi}
\ee
where $U(\phi)$ is the potential of $\phi$ and derivative is taken with 
respect to $\phi$. This anzats gives reasonable results only for harmonic
potential but in all other cases this approximation is not satisfactory. 
A better approximation has been derived in refs.~\cite{dolgov95,dolgov99}. 
One starts with exact quantum operator equation of motion for the field
$\phi$ and some other fields $\chi$ that are coupled to $\phi$. The 
production of the latter by oscillations of $\phi$ results in a damping 
term in the equation of motion for $\phi$. 
As an example let us consider a simple case of scalar $\chi$ with trilinear
coupling $f\phi \chi^2$. The corresponding equations of motion have the
form (expansion neglected for simplicity):
\be
\ddot{\phi} -\Delta \phi + V^{'}(\phi) &=& f \chi^2 ,
\label{eqvp}\\
\partial^2 \chi +m^2_\chi\, \chi &=& 2 f \phi \chi.
\label{eqchi}
\ee
The next step is to make quantum averaging of these equations in the
presence of classical field $\phi_c (t)$ (in what follows we omit sub-c and
neglect the mass of $\chi$). This can be easily done in one-loop
approximation (some subtleties related to renormalization of mass and 
coupling constants are discussed in ref.~\cite{dolgov99}) and one comes to 
the equation that contains only the field $\phi$ and accounts for the
backreaction from the production of the quanta of $\chi$:
\be
 \ddot{\phi} + V'(\phi)   = 
\frac{f^2}{4 \pi^2} \int_0^{t-t_{in}} \frac{d \tau}{\tau} \, \phi(t-\tau)~,
\label{scalres}
\ee
where $t_{in}$ is an initial time, when the particle production was switched 
on (it is assumed that $t>t_{in}$). The term in the r.h.s. that describes
the influence of the particle production is non-local in time as one should 
have expected because the impact of the produced particle on the evolution 
of $\phi$ depends upon all the previous history. To use this equation for
realistic calculations one has to make proper renormalization procedure. It
is described in detail in ref.~\cite{dolgov99}. The coupling to fermions as
well as quartic coupling $\lambda' \phi^2 \chi^2$ are also considered 
in that paper.
Similar one-loop approach was used in ref.~\cite{baacke98} but no 
self-contained equation for $\phi$ was derived there.

Both effects of cosmological expansion and of damping of $\phi$ due to
particle production can be easily incorporated into imaginary time method. 
This is especially simple in the case of fast oscillations and slow decrease
of the amplitude of $\phi$. In this case the results obtained above
practically do not change. One should only substitute there $\phi_0 (t)$
and to determine the law of the evolution of the latter from the energy
balance equation~(\ref{dotrho}) or, more accurately, from eq.~(\ref{scalres}).

One more phenomenon deserves a comment here. As we have already mentioned,
production of bosons may be strongly amplified due to the presence of
the earlier produced bosons in the same final state. In classical language 
this effect is
described by the parametric resonance in the equation of motion of the
produced particles, while in quantum language it is the so called stimulated
emission well known in laser physics. When the amplitude of the driving field
$\phi$ drops below a certain value, the resonance would not be excited and
the rest of $\phi$ would decay slowly. If the mass of $\phi$ is non-zero,
this field would behave as non-relativistic matter and its cosmological
energy density would drop as $1/a^3$. On the other hand, the produced
particles are mostly relativistic with energy density decreasing as
$1/a^4$. Thus for a sufficiently slow decay rate of $\phi$ the latter
may dominate the cosmological energy density once again,
when previously produced particles are red-shifted away. This would result 
in a low second reheating temperature,
much lower than in parametric resonance scenario. On the other hand, the
phenomenon of stimulated emission persists in perturbation theory even with
a very small amplitude of $\phi$. Possibly even in this limit the production
is not very fast as well, because the width of the band is quite narrow and
the produced bosons are quickly pushed away from the band due to cosmological 
red-shift and collisions. More detailed consideration is desirable here.

\section{Conclusion. \label{s:concl}}

It is demonstrated that imaginary time method very well describes
particle production by scalar field. It is very simple technically and
permits to obtain physically transparent results. The calculations here
were done for a particular case of periodic or quasiperiodic oscillations
of the field but, as shows the experience with production of $e^+e^-$-pairs
by electric field (for a review see e.g. third paper in
ref.\cite{marinov72}),
the method also works well in the opposite case of short pulse fields. 
The method is applicable in the quasiclassical limit. In the opposite case
perturbation theory is applicable and hence one can obtain simple and
accurate (semi)analytical estimates practically in all parameter range.

The results of calculations in the quasiclassical limit are in a good
agreement with subsequent numerical ones\,\cite{baacke98,green99}.
An important difference between the latter papers and the initial
one\,\cite{dolgov90} lays in the interpretation of the results. According
to all these papers the occupation numbers of the produced particles quickly
approaches unity but, in contrast to refs.\cite{baacke98,green99}, it is
argued in the paper\,\cite{dolgov90} that the total production rate is 
nevertheless suppressed in comparison to perturbation theory and
the production of fermions by the inflaton with Yukawa coupling to 
fermions is always weak. This conclusion is verified above.
As is shown in this paper, the occupation numbers may quickly
reach unity both in perturbation theory and in non-perturbative case. 
Still even the production rate of particles obeying Boltzmann statistics 
is very weak to ensure fast (pre,re)heating. In the case of fermion 
production the rate is evidently
much weaker because the production must stop when the occupation number
reaches unity and to continue the process the produced fermions should be 
eliminated from the band. As is argued in sec.~\ref{ss:small}, the 
non-perturbative effects can only diminish the production rate.

The bosonic case is opposite: more bosons are in the final state, the
faster is production. Thus even in perturbation regime the boson production
can be strongly amplified because their occupation number may reach unity
in much shorter time than $1/\Gamma$ and the energy may be transferred from 
the inflaton to the produced bosons much faster than 
is given by the original perturbative
estimates\,\cite{dolgov82}, where the effect of stimulated emission was not 
taken into account. Of course to realize this regime the band should not be 
destroyed by expansion and scattering, as argued in ref.~\cite{dolgov90}.

To summarize, we have shown that perturbation theory gives a good estimate 
of production of light fermions and bosons if Fermi exclusion principle 
or stimulated emission respectively are taken into account. The formally
calculated production rate in perturbation theory is always larger than
the non-perturbative one, at least in 
the simple cases that we have considered.
So the results of perturbation theory may be used as upper bounds for 
production rates. Moreover, perturbation theory helps to understand physical
meaning of the obtained results and to interpret them correctly.

In many realistic cases (e.g. for large $g\phi_0$ or $m_0$)
perturbation theory is not applicable and to 
calculate the real production rate (not just an upper bound) one has to 
make more involved 
non-perturbative calculations. In quasiclassical (anti-perturbative)
limit imaginary time method permits to obtain accurate and simple
results and to avoid complicated numerical procedure
\section*{Acknowledgments}
\addcontentsline{toc}{section}{\numberline{}Acknowledgments}

I am grateful to S. Hansen and A. Vainshtein for critical comments on the 
manuscript.

\section*{References}
\addcontentsline{toc}{section}{\numberline{}References}

\end{document}